\documentclass{aa}  

\usepackage{graphicx}
\usepackage{amssymb}
\usepackage{mathtools}
\usepackage{amsmath}
\usepackage{xcolor}
\usepackage{adjustbox}
\usepackage{ulem}
\usepackage{txfonts}
\usepackage{fancyvrb}
\newcommand{\ep}{E_{\mathrm{peak}}}
\newcommand{\eb}{E_{\mathrm{break}}}

\newcommand{\epb}{E_{\mathrm{peak, Band}}}
\newcommand{\eps}{E_{\mathrm{peak,2SBPL}}}
\newcommand{\ebs}{E_{\mathrm{break, 2SBPL}}}
\newcommand{\bs}{\beta_{\mathrm{2SBPL}}}
\newcommand{\au}{\alpha_{\mathrm{1,2SBPL}}}
\newcommand{\ad}{\alpha_{\mathrm{2,2SBPL}}}

\newcommand{\ab}{\alpha_{\mathrm{Band}}}
\newcommand{\bb}{\beta_{\mathrm{Band}}}

\newcommand{\chib}{\chi^2_{\mathrm{r,Band}}}
\newcommand{\chis}{\chi^2_{\mathrm{r,2SBPL}}}

\newcommand{\er}{R_{\mathrm{E}}}

\begin{document}

   \title{The slope of the low energy spectrum of Gamma-Ray Burst prompt emission}

   \subtitle{}


   \author{
   M. Toffano\inst{1}\thanks{E--mail: mattia.toffano@inaf.it}
\and G.~Ghirlanda\inst{2,3} 
\and L. Nava\inst{2,8,9}
\and G. Ghisellini\inst{2}
\and M. E. Ravasio\inst{4}
\and G. Oganesyan\inst{5,6,7}
}

\authorrunning{M. Toffano et al.}

   \institute{
$^1$ Università degli Studi dell'Insubria, Via Valleggio 11, 22100, Como, Italy\\
$^2$ INAF -- Osservatorio Astronomico di Brera, Via E. Bianchi 46, I-23807 Merate, Italy\\
$^3$ INFN -- Sezione di Milano-Bicocca, Piazza della Scienza 3, I-20126 Milano, Italy \\
$^4$ Università degli Studi di Milano-Bicocca, Dip. di Fisica ``G. Occhialini'', Piazza della Scienza 3, I-20126 Milano, \\
$^5$ Gran Sasso Science Institute, Viale F. Crispi 7, I-67100, L’Aquila (AQ), Italy\\
$^6$ INFN - Laboratori Nazionali del Gran Sasso, I-67100, L’Aquila (AQ), Italy\\
$^7$ INAF - Osservatorio Astronomico d’Abruzzo, Via M. Maggini snc, I-64100 Teramo, Italy\\
$^8$ INFN -- Sezione di Trieste, via Valerio 2, I-34149 Trieste, Italy \\
$^9$ Institute for Fundamental Physics of the Universe (IFPU), I-34151 Trieste, Italy}

\date{Received xxx / Accepted: xxx}

 
\abstract{

Gamma-ray Bursts (GRBs) prompt emission spectra are often fitted with the empirical 
``Band" function,  namely two power laws smoothly connected. 
The typical slope of  the low energy (sub-MeV) power law is $\ab\simeq -1$.
In a small fraction  of long GRBs this power law splits into two components such 
that the spectrum presents, in addition to the typical $\sim$MeV $\nu F_{\nu}$ peak, a 
break  at the order of a few keV or hundreds keV.   
The typical power law slopes below and above the break are --0.6 and --1.5 respectively. 
If the break is a common feature, the value of $\ab$ could be an ``average'' of 
the spectral slopes below and above the break in GRBs fitted with Band function. 
We analyze the spectra of 27 (9) bright long (short) GRBs 
detected by the {\it Fermi} satellite finding a low energy break between 
80 keV and 280 keV in 12 long GRBs, 
but in none of the short events.
Through spectral simulations we show that if the break is moved closer (farther) to the peak energy a relatively harder (softer) $\ab$ is found by fitting the simulated spectra with the Band function.
The hard average slope $\ab\simeq-0.38$ found in short GRBs suggests that the break 
is close to the peak energy. We show that for 15 long GRBs best fitted by the 
Band function only, the break could be present, but it is not identifiable in 
the {\it Fermi}/GBM spectrum, because either at low energies, close to the detector 
limit for relatively soft $\ab\lesssim-1$, or in the proximity of the energy peak 
for relatively hard $\ab\gtrsim-1$.
A spectrum with two breaks could be typical of GRB prompt emission, 
though hard to identify with  current detectors.
Instrumental design such that conceived for the  THESEUS space mission, 
extending from 0.3 keV to several MeV and featuring a larger effective 
area with respect to Fermi/GBM, can reveal a larger fraction of GRBs with 
a spectral energy break. 
}
   

\maketitle
%

\section{Introduction}

The prompt emission spectra of Gamma-Ray Bursts (GRBs) are usually described by empirical functions.
The most commonly used is the Band function  \citep{Band1993}, which consists of two smoothly-connected power-laws $N(E)\propto E^{\alpha_{\mathrm{Band}}}$ and $N(E)\propto E^{\beta_{\mathrm{Band}}}$ describing the photon spectrum at low and high energies, respectively. Spectral studies  of long  GRBs (LGRBs; i.e. observed duration $>2$ seconds, \citealt{Kouveliotou1993})  detected by the Burst And Transient Source Experiment 
(BATSE)  on board the {\it Compton Gamma Ray Observatory} (CGRO) showed that, on average, $\ab\sim-1$, $\bb\sim -2.3$ and the $E^2 N(E)$ spectral energy distribution peaks at $\ep\sim 300$ keV \citep{kaneko06}. Similar results were obtained by spectral studies of GRBs detected by the \textit{Fermi} Gamma--ray Burst Monitor (GBM;  \citealt{gruber14, vonkienlin20, nava11a}) and the 
{\it Beppo}SAX Gamma--Ray Burst Monitor (GRBM; \citealt{frontera09}). Short GRBs (SGRBs) have on average a harder $\ab\sim -0.6$ \citep{ghirlanda04,ghirlanda09} and a larger observed peak energy.\footnote{Accounting for the different redshift distributions, the peak energy of short and long become similar \citep{ghirlanda15,nava12}}   

Recently, in a sizable fraction of LGRBs detected simultaneously by the Burst Alert Telescope (BAT) and the X-Ray Telescope (XRT) aboard the {\it Neil Gehrels Swift Observatory} (hereafter {\it Swift}), and in the brightest GRBs detected by 
{\it Fermi}/GBM it was found that (a) a low energy spectral break is present, 
between $\sim 1$ and few hundreds keV,  and (b) the slopes of the power-law below and above this break are 
consistent with the values expected from synchrotron emission \citep{oganesyan17, oganesyan18,ravasio18,ravasio19}. 
The consistency of GRB spectra with synchrotron emission was also proved by direct fits with a synchrotron model \citep{oganesyan19,Chand2019,burgess20, ronchi20}. 
The interpretation of these results within the leptonic synchrotron scenario is quite challenging, requiring a small 
magnetic field ($\sim$10 G) and a large dissipation radius ($10^{16-17}$ cm) \citep{Kumar2008,daigne11,Beniamini2013,oganesyan19,ronchi20}. 
Synchrotron emission from protons has been proposed as a possible solution \citep{ghisellini20}, but the matter is still under debate \citep{florou21}. 

While the presence of a low energy spectral break has been identified in a subsample of GRBs, a sizable fraction of GRB spectra are well fitted by single break functions where the break energy corresponds to $E_{\rm peak}$. If the presence of an additional break located at lower energies is a common feature in GRB spectra, its non detection could be due to several factors. For spectra with a low signal-to-noise  (S/N) ratio  the identification of a change in slope might be difficult. Also, if the break is either close to the peak energy or below the low energy edge of the instrument sensitivity, a spectral fit would be unable to identify the break.  \cite{burgess15} studied how the Band function fits spectra simulated assuming a synchrotron or a synchrotron plus black--body model.

In this work we investigate the possibility that the low energy spectral break is a common feature in GRB spectra, and that the average spectral index $\ab\sim-1$ is an average value between the two power-law segments below and above the break energy. To this aim we 
analyze bright long and short GRBs detected by {\it Fermi} (\S2,3) in search for a statistically significant evidence of the low energy spectral break. 
Through spectral simulations we study how the value of $\ab$ becomes relatively harder (softer) by moving the break closer to  (farther from) the peak energy (\S4). 
We derive (\S4.2) limits on the possible location of the break energy in GRBs whose spectrum is best fitted by the Band function (i.e. with no evidence of a break). We discuss and summarize our results in \S5.

\section{The sample}
To perform our investigation, we need first to characterize the low energy part of GRB spectra, searching for possible spectral breaks like those recently reported in a number of events. To this aim, we select GRBs detected by the GBM (8 keV--40 MeV; \citealt{Meegan2009}) and apply selection criteria that maximize the probability to i) perform reliable spectral analysis and ii) identify the presence of a spectral break at low energies, well distinguished from the spectral peak energy.
This can be achieved by selecting GRBs with large fluence and large $\ep$ 
values, as they have better chances to display the 
low energy break $\eb$ within the sensitivity energy range of the GBM, given that their typical separation is $\eb/\ep \sim$ 0.1 \citep{ravasio19}. 

For these reasons, from the online {\it Fermi} catalog\footnote{\url{https://heasarc.gsfc.nasa.gov/W3Browse/fermi/fermigbrst.html}}, 
which contains 2669 GRBs up to April 2020, we select LGRBs\footnote{
GRBs are classified into long and short based on the values of $T_{90}$ reported in the 
\textit{Fermi} on-line catalog, with a separation at 2 seconds.} with fluence  $>10^{-4} \, \mathrm{erg \, cm^{-2} }$ 
(integrated over the 10--1000 keV energy range) and observed peak energy $\ep$ $>300$\,keV, and SGRBs with fluence $>5\times 10^{-6} \, 
\mathrm{erg \, cm^{-2} }$ and $\ep>300$\,keV. 
This selection is based on the values of fluence and peak energy reported in the catalog and 
corresponding to the fit of the time-integrated spectrum with the Band function.
We excluded GRB\,090902B since the presence of a power-law component 
(in addition to a multi--temperature black body - \citealt{Ryde2010,Liu2011,Peer2012}), 
hampers the identification of the spectral break at low energies. 
Our selection results in 27 long and 9 short GRBs.

\section{Spectral analysis}

The $Fermi$/GBM consists of 12 sodium iodide (NaI - 8--1000 keV) and 2 bismuth germanate (BGO - 0.2--40 MeV) detectors. For each burst we analyzed the data from two NaI detectors and one BGO, selected as those with the smallest GRB position angle with respect to the detector normal.

We used the public tool 
\texttt{gtburst}\footnote{\url{https://fermi.gsfc.nasa.gov/ssc/data/analysis/scitools/gtburst.html}} to  retrieve spectral CSPEC data from the {\it Fermi} Data Server\footnote{\url{https://fermi.gsfc.nasa.gov/ssc/data/access/}}. We extract the time integrated spectrum over the time interval 
$T_{90}$ 
reported in the {\it Fermi} catalog. In the case of GRB~130427A and GRB~160625B, which have a multi-peaked, 
long-duration light curve, we consider the brightest portion of the light 
curve by selecting the time intervals $3.0-15$ s and $188-210$ s, respectively.\footnote{ A detailed analysis of the first peak ($0 - 2.5$s) of GRB~130427A has been  performed in \cite{preece14}}.
The background spectrum has been estimated by selecting two time intervals before and after 
(far from) the GRB emission episode. The latest detector response matrices for each event were 
obtained through \texttt{gtburst}.

Standard energy selections\footnote{\url{https://fermi.gsfc.nasa.gov/ssc/data/analysis/scitools/}} for the analyzed spectra were applied (8--900 keV for NaIs and 0.3--40 MeV for BGO) and the 30-40 keV range was excluded to avoid contamination of inaccurate modeling  of the Iodine K-edge line at 33.17 keV in the response files. The spectral data of the two NaI and one BGO  were fitted together accounting for an inter-calibration constant parameter. 

The extracted spectra were analyzed with \texttt{XSPEC}\footnote{\url{https://heasarc.gsfc.nasa.gov/xanadu/xspec/}} (v12.10.1).
All spectra were fitted with two different spectral models: 
\begin{itemize}
\item the Band function \citep{Band1993}, often used to fit GRB spectra \citep{preece98, kaneko06, ghirlanda02, nava11,sakamoto11, gruber14, yu16, lien16, goldstein12, frontera00}:
\begin{equation} \label{eq:band}
N(E)\propto\begin{cases} 
              E^{\alpha}e^{-E/E_0},  & \mathrm{if} \, E\leq (\alpha-\beta)E_0 \\
              [(\alpha-\beta)E_0]^{(\alpha-\beta)}E^{\beta}e^{(\beta-\alpha)}, & \mathrm{if} \, E> (\alpha-\beta)E_0 
               
               
            \end{cases}
\end{equation}

\noindent where $\alpha$ and $\beta$ are the photon indices of the power-laws below and above, respectively, the e-folding energy $E_0$. $N(E)$ has units of ph cm$^{-2}$ s$^{-1}$ keV$^{-1}$. For $\beta<-2<\alpha$ the spectrum in the $E^2N(E)$ representation peaks at $E_{\rm peak}=E_0(\alpha+2)$. 

\item the double smoothly-broken power law function (2SBPL hereafter,  \citealt{ravasio18})

\begin{eqnarray}\label{eq:2SBPL}
N(E) 
&\propto& E^{\alpha_1}_{\mathrm{break}}\cdot
 \left\{ \left[ \left( \frac{E}{\eb} \right)^{-\alpha_1 n_1} 
+ \left(\frac{E}{\eb}\right)^{-\alpha_2 n_1}\right]^{n_2/n_1} \right. \nonumber \\
&+& \left. \left( {\frac{E}{E_j}} \right)^{-\beta n_2} 
\cdot 
 \left[  \left( \frac{E_j}{\eb} \right)^{-\alpha_1 n_1} + 
 \left( \frac{E_j}{\eb} \right)^{-\alpha_2 n_1} \right]^{n_2/n_1} \right\}^{-1/n_2} 
\end{eqnarray}
where 
\begin{equation}
E_j = \ep\cdot\left(- \frac{\alpha_2 +2}{\beta +2}\right)^{1/[(\beta-\alpha_2)n_2]}, 
\end{equation}
and $\ep$ and $\eb$ are the peak of the $\nu F_{\nu}$ spectrum and the break energy, respectively. The parameters $\alpha_1$ and $\alpha_2$ are the power-law indices below and above the break energy, $\beta$ is the power-law index above the peak energy. The parameters $n_1$ and $n_2$ set the sharpness of the curvature around $\eb$ and $\ep$, respectively. Following \cite{ravasio19}, we assumed $n_1=n_2=2$. 

\end{itemize}

In the following, in order to distinguish the spectral parameters of these two fitting functions, we call $\ab$ and $\bb$ the photon indices of the Band function and  $\au$ and $\ad$ the photon indices of the 2SBPL below the peak energy and $\beta_{\mathrm{2SBPL}}$ above it. 
 
The large number of counts of the extracted spectra allow us to fit the spectra and search for the best fit parameters by minimizing the $\chi^2$ statistics. 
We adopt the Akaike Information Criterion (AIC - \citealt{akaike74}) to compare the fits obtained with the 2SBPL and Band functions and choose the best one. We recall that $\mathrm{AIC} = 2k - 2\ln(\hat{L})$, where $k$ is the number of free parameters in the model and $\hat{L}$ is the maximum value of the likelihood function $L$ obtained by varying the free parameters. For Gaussian-distributed variables $\chi^2\propto -2\ln(L)$.  If $\Delta{\mathrm{AIC}}=\mathrm{AIC}_{\rm Band}-\mathrm{AIC}_{\rm 2SBPL}\ge6$, the Band fit has $\lesssim 5\%$  probability of describing the observed spectrum better than the 2SBPL function \citep{akaike74}: in such case, we consider the 2SBPL a better fit and thus consider the presence of a break as statistically significant at the 95\% confidence level.

\begin{table*}[htbp]
\renewcommand*{\arraystretch}{1.4}
    \centering
    		\vspace*{-0.7cm} 
    
    \begin{adjustbox}{angle=90}
    \begin{tabular}{c|c|c|c|c|c|c|c|c|c|c|c|c}
    \hline
    \hline
    Name & $\ab$ & $\epb$ & $\bb$ &$\au$ & $\ad$ & $\ebs$& $\eps$ & $\bs$ & $\chib$ & $\chis$& $\mathrm{AIC_{Band}}$ & $\mathrm{AIC_{2SBPL}}$\\ 
    
    &&&&&&&&&&\\


     100724(029) & $-0.71_{-0.01}^{+0.01}$ & $339_{-12}^{+12}$ & $-2.09_{-0.03}^{+0.03}$ & $-0.80_{-0.02}^{+0.01}$ & $-1.90^{+0.05}$
    & $168_{-13}^{+23}$ & $345_{-226}^{+380}$ & $-2.35_{0.11}^{+0.10}$ & 0.99 & 0.92 & 463 & 441   \\

       100918(863) & $-0.74_{-0.04}^{+0.05}$ & $ 406_{-49}^{+51}$ & $ -2.47_{-0.17}^{+0.12}$ & $-0.74_{-0.05}^{+0.07}$ &
      $<-1.77$
      & $146_{-34}^{+20}$ & $523_{-124}^{+159}$ & $-3.17^{+0.36}_{-0.5}$ & 0.94& 0.89& 221& 213 \\
       130427(324) & $-0.66_{-0.01}^{+0.01}$ & $852_{-6.7}^{+6.7}$ & $-3.27_{-0.03}^{+0.03}$ & $-0.63_{-0.01}^{+0.01}$ & $-1.67_{-0.03}^{+0.03}$ & $224_{-10}^{+10}$&$992_{-12}^{+12}$ & 
      $-3.7_{-0.04}^{+0.04}$ & 6.01 & 2.83 &942 &812\\

       131014(215) & $-0.21_{-0.01}^{+0.01}$& $306_{-4}^{+4}$& $-2.72_{-0.02}^{+0.02}$ & $-0.33_{-0.01}^{+0.01}$ & $-1.8_{-0.04}^{+0.04}$& $124_{-6}^{+6}$ & $386_{-15}^{+14}$ & 
      $-3.48_{-0.10}^{+0.09}$ & 2.17 & 1.35 &642 &620 \\
        131028(076) & $-0.66_{-0.01}^{+0.01}$ & $860_{-25}^{+28}$ & $-3.34_{-0.18}^{+0.14}$ & $-0.65_{-0.02}^{+0.02}$ & $-1.68_{-0.07}^{+0.07}$ &$249_{-23}^{+23}$ & $991_{-56}^{+64}$& $-3.74_{-0.2}^{+0.17}$ & 1.26 & 1.14 &438&401\\
           150510(139)& $-1.01_{-0.01}^{+0.01}$ & $1172_{-52}^{+118}$ &
      $<-3.88$
      & $-0.86_{-0.04}^{+0.05}$ & $-1.52_{-0.11}^{+0.10}$ & $173_{-35}^{+42}$ & $1776_{197}^{+221}$ & $-4.90_{-0.89}^{+0.61}$ & 1.07 & 0.97 &374 & 342\\
 160509(374) & $-0.89_{-0.04}^{+0.04}$ & $468_{-53}^{+60}$ & $-2.73_{-0.14}^{+0.13}$ & $-0.63_{-0.12}^{+0.10}$ & $-1.66_{-0.07}^{+0.08}$& $80_{-17}^{+22}$ & $2071_{-545}^{+635}$ & $-2.82_{-0.12}^{+0.14}$ & 1.01 &0.95 & 484 & 460\\
      160625(945) & $-0.54_{-0.01}^{+0.01}$ & $362_{-6}^{+6}$ & $-2.26_{-0.01}^{+0.01}$& $-0.55_{-0.01}^{+0.01}$ & $-1.71_{-0.02}^{+0.02}$ & $120_{-3}^{+3}$ & $684_{-20}^{+21}$ & 
      $-2.75_{-0.03}^{+0.03}$ & 4.73 & 1.64 &1607 &568\\
       160821(857) & $-0.95_{-0.01}^{+0.01}$ & $836_{-17}^{+17}$ & $-2.22_{-0.02}^{+0.02}$& $-0.86_{-0.03}^{+0.04}$ & $-1.56_{-0.09}^{+0.08}$ & $150_{-24}^{+23}$ & $1362_{-66}^{+74}$  &  
      $-2.6_{-0.09}^{+0.11}$ & 2.06 & 1.53 &932 &699  \\
      170409(112) & $-0.79_{-0.01}^{+0.01}$ & $839_{-25}^{+27}$ & $-2.58_{-0.05}^{+0.04}$ & $-0.79_{-0.01}^{+0.01}$ & $-1.81_{-0.04}^{+0.04}$ & $278_{-15}^{+15}$ & $1292_{-105}^{+118}$ &      $-3.48_{-0.16}^{+0.19}$ & 1.53 & 1.27 &700 &588\\
      171227(000)   & $-0.81_{-0.01}^{+0.01}$ & $737_{-31}^{+33}$ & $-2.53_{-0.04}^{+0.04}$ & $-0.63_{-0.05}^{+0.05}$ & $-1.49_{-0.07}^{+0.07}$ & $112_{-22}^{+24}$ & $968_{-59}^{+70}$&$-2.86_{-0.07}^{+0.09}$ & 1.37&0.95 & 320& 227 \\
      
       180720(598) & $-1.04_{-0.01}^{+0.01}$ & $472_{-14}^{+15}$ & $-2.37_{-0.06}^{+0.05}$ & $0.97_{-0.04}^{+0.05}$ & $1.78_{-0.05}^{+0.06}$ & $121_{-17}^{+15}$ & $951_{-106}^{+135}$ & 
      $-2.99_{-0.13}^{+0.16}$ & 1.87 & 1.49 &853 &685\\
   

\hline
         \end{tabular}

\end{adjustbox}

    \caption{Results of the fits for GRBs best fitted by a 2SBPL. The parameters obtained by fitting a Band function are also shown for comparison. 
    Analyzed GRBs  are listed in column 1 (in parenthesis, the trigger name according to the \textit{Fermi} catalog).
    The parameters concerning the Band function (columns 2 -- 4) and the 2SBPL function
    (columns 5--9) are defined in Eq.~\ref{eq:band} and Eq.~\ref{eq:2SBPL}, respectively. 
    The reduced $\chi^2$ statistics ($\chi^2_r$) and the AIC for each fit are shown in 
    columns 10 - 13. 
    Both errors and upper limits are calculated at $1\sigma$ confidence level.
    Energies are expressed in keV.}
 \label{tab:fit_params_w/_break}
\end{table*}

\begin{table*}[htbp]
\renewcommand*{\arraystretch}{1.4}
    \centering
    \begin{tabular}{c|c|c|c|c|c}
       \hline
    \hline
    Name & $\ab$ & $\epb$ & $\bb$ & $\chib$ & $\mathrm{AIC_{Band}}$ \\
    &&&&&\\

    090323(002) & $-1.08_{-0.02}^{+0.03}$  & $340_{-30}^{+34}$ &  $-2.47_{-0.33}^{+0.17}$ & 0.90 &420 \\


      090926(181) & $-0.70_{-0.01}^{+0.01}$ & $291_{-5}^{+5}$ & $-2.64_{-0.05}^{+0.04}$ &1.50 &861 \\
     100414(097) & $-0.49_{-0.02}^{+0.02}$ & $561_{-19}^{+21}$ & 
    $<-4.7$&1.14 & 518 \\
   101123(952)& $-0.95_{-0.02}^{+0.02}$  & $448_{-40}^{+45}$ & $-2.31_{-0.15}^{+0.11}$&0.88 & 307 \\
     120526(303) & $-0.81_{-0.03}^{+0.03}$ &  $752_{-62}^{+68}$ & $-3.35_{-0.62}^{+0.33}$  &0.90& 215 \\

      120624(933) & $-0.95_{-0.03}^{+0.03}$ & $580_{-65}^{+75}$  & $-2.30_{-0.18}^{+0.12}$&1.00 & 464 \\

     120711(115) & $-0.93_{-0.01}^{+0.01}$  & $1108_{-79}^{+65}$ & $-3.26_{-0.20}^{+0.15}$&0.94 & 223 \\
 130306(991) & $-0.81_{-0.17}^{+0.21}$ & $211_{-51}^{+64}$  & 
     $<-2.78$&0.59 & 263 \\
                        
      130504(978) & $-1.22_{-0.02}^{+0.02}$  & $656_{-74}^{+86}$& $-2.46_{-0.17}^{+0.13}$& 0.97 & 229 \\

     130606(497)& $-1.05_{-0.01}^{+0.01}$  & $329_{-14}^{+16}$& $ -2.11_{-0.02}^{+0.02}$&1.52 & 692 \\

     140206(275) & $-1.33_{-0.02}^{+0.02}$  & $309_{-29}^{+33}$& $-2.19_{-0.11}^{+0.09}$&0.98 & 432 \\
160905(471) & $-0.86_{-0.02}^{+0.02}$  & $905_{-61}^{+71}$ & $-3.10_{-0.75}^{+0.29}$&1.16 & 534 \\
    170210(116) & $-1.08_{-0.02}^{+0.02}$  & $500_{-47}^{+54}$& $-2.36_{-0.17}^{+0.12}$&0.91 & 320 \\

     170214(649) & $-0.82_{-0.03}^{+0.03}$  & $370_{-22}^{+25}$ & $-2.48_{-0.12}^{+0.10}$&1.15 & 532 \\

     170527(480) & $-1.03_{-0.02}^{+0.02}$ & $823_{-52}^{+56}$  & 
     $<-4.21$&1.09 & 502 \\

     %

    \hline

    \end{tabular}
    \caption{Results of the fits for GRB best fitted by a Band function. 
    Analysed GRBs  are listed in column 1 (in parenthesis, the trigger name according to the \textit{Fermi} catalog).
    The parameters regarding the Band function (columns 2 - 4) are defined in Eq.~\ref{eq:band}.  
    The reduced $\chi^2$ statistics ($\chi^2_r$) and the AIC for each fit are shown in columns 
    5 -- 6.
    Both errors and upper limits are calculated at $1\sigma$ confidence level. Energies are expressed in keV. }
    \label{tab:fit_params_w/o_break}
\end{table*}

\begin{table*}[htbp]
\renewcommand*{\arraystretch}{1.4}
    \centering
    \begin{tabular}{c|c|c|c|c|c}
    \hline
    \hline
 Name & $\ab$ & $\epb$ & $\bb$ & $\chib$ & $\mathrm{AIC_{Band}}$\\ 
    &&&&&\\

    090227(772) & $-0.48_{-0.17}^{+0.19}$ & $1830_{-136}^{+108}$ & $-3.05_{-0.26}^{+0.19}$ & 1.01 & 336\\
 090228(204) & $-0.5_{-0.02}^{+0.02}$ & $621^{+50}_{-48}$ & 
$<-4.41$
& 1.12 &482\\
 111222(619) & $-0.28_{-0.07}^{+0.08}$ & $616_{-58}^{+64}$ & 
$<-4.04$& 1.21 & 358\\
          
 130504(314) & $-0.28_{-0.05}^{+0.04}$ & $990_{-72}^{+75}$ &
$<-4.52$&  1.04 &337 \\
 130701(761) & $-0.47_{-0.06}^{+0.06}$ & $869_{-98}^{+107}$ & 
$<-4.56$ & 0.98 & 443\\
          
   150819(440) & $-1.04_{-0.03}^{+0.03}$ & $400_{-22}^{+45}$ & 
    $<-4.09$ & 0.95 & 314 \\
         
 170127(067) & $ +0.3_{-0.13}^{+0.14}$ & $768_{-62}^{+64}$ & $-4.05_{-1.01}^{+0.51}$ & 1.00 &176\\

 170206(453) & $-0.12_{-0.05}^{+0.05}$ & $274_{-17}^{+17}$ & $-2.76_{-0.11}^{+0.10}$ & 1.00 &442 \\

 170222(209) & $-0.56_{-0.07}^{+0.08}$ & $700_{-89}^{+100}$ & 
$<-4.41$& 0.89 &312 \\

   \hline      
    \end{tabular}
    \caption{Results of the fits performed using the Band function for the  SGRBs in our sample. Analyzed GRBs  are listed in column 1 (in parenthesis, the trigger name according to the \textit{Fermi} catalog).
    The parameters regarding the Band function (columns 2 -- 4) are defined in Eq.~\ref{eq:band}. 
    The reduced $\chi^2$ statistics ($\chi^2_r$) and the AIC for each fit are shown in 
    columns 5 -- 6.
    Both errors and upper limits are calculated at $1\sigma$ confidence level. Energies are expressed in keV.}
    \label{tab:fit_params_short}
\end{table*}

\subsection{Fit results: best fit model}\label{sec:results_data}
The fit results for LGRBs are reported in Table~\ref{tab:fit_params_w/_break} and \ref{tab:fit_params_w/o_break}. The fit results for SGRBs are shown in Table~\ref{tab:fit_params_short}. The errors on the parameters represent the $1\sigma$ confidence\footnote{through the \texttt{error} method built in \texttt{XSPEC}}. 

We find that: 
\begin{itemize}

\item twelve (out of 27) LGRBs have a low energy break, i.e. their spectra are 
best fitted by the 2SBPL function ($\Delta{\mathrm{AIC}} \ge 6$). 
The spectral parameters are reported in Table~\ref{tab:fit_params_w/_break};

\item the remaining fifteen LGRBs are well fitted by the Band function and, according 
to the AIC criterion, there is no improvement using the 2SBPL function. 
Their spectral parameters are reported in Table~\ref{tab:fit_params_w/o_break};

\item all SGRBs are well fitted by the Band function. In six SGRBs 
we could only derive an upper limit on $\bb$,
indicating that also a cutoff power-law function could be a good fit to the spectra 
(see e.g. \citealt{ghirlanda04}).
\end{itemize}

\begin{figure}[htbp]
    \begin{center}
       \hspace*{-0.2cm}
    \includegraphics[scale=0.54]{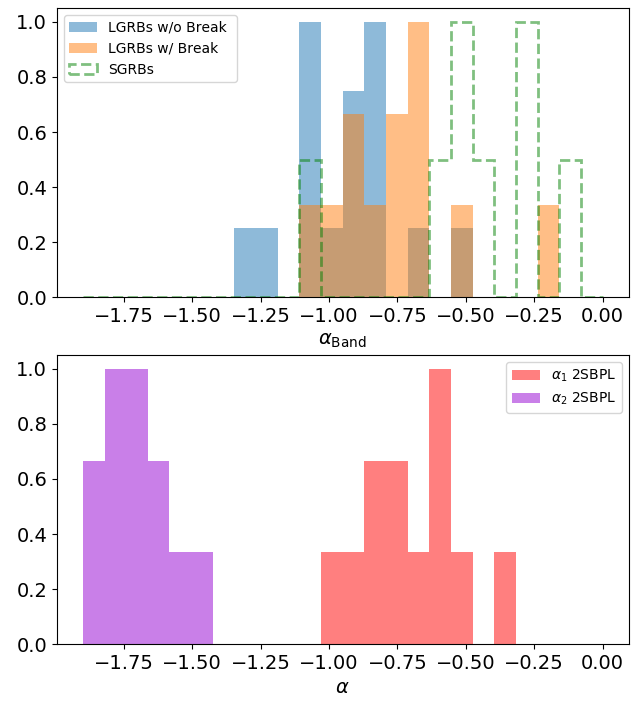}
    \caption{Top:  distributions of $\ab$ for SGRBs (green) and for both  LGRBs 
    with and without the low energy spectral break (orange and blue histogram). 
    Bottom:  distributions of $\au$ and $\ad$ of the 12 LGRBs best 
    fitted by the 2SBPL (i.e. with the low energy spectral break). Distributions are normalized to their peak values.}
    \label{fig:a1_a2_ab_dist} 
    \end{center}
\end{figure}

In the LGRB 160509A, we  find a well constrained $\eb\simeq 80$\,keV but the peak energy of the 2SBPL is 
undetermined by fitting the GBM data. 
Only in this case, we exploited the LAT Low Energy (LLE) data to better constrain the high-energy 
index $\beta$ and thus $\ep$.
With \texttt{gtburst} we extracted the time-integrated spectrum from the LLE data \footnote{\url{https://heasarc.gsfc.nasa.gov/W3Browse/fermi/fermille.html}} 
and performed a joint GBM-LLE spectral fit over the 10 keV-- 300 MeV energy range. 
Assuming an intercalibration  
normalization factor between LAT-LLE and NaI detectors of 1, we obtained an estimate of $\ep\simeq 2071$\,keV for GRB~160509A (Table~\ref{tab:fit_params_w/_break}).

\subsection{Fit results: spectral indices below $E_{\rm peak}$}
Figure~\ref{fig:a1_a2_ab_dist} (top panel) shows the distribution of the spectral index $\ab$ for the entire sample. The blue histogram corresponds to LGRBs without the break and the green dashed histogram is for SGRBs (all without a break). For comparison it is also shown the distribution of $\ab$ for the 12 LGRBs whose spectrum is better fitted by the 2SBPL.

\begin{table}[htbp]

\centering
    \begin{tabular}{c|c|c|c|c}
    \hline
    \hline
          GRB type & $N$ & $\langle \ab \rangle$ &$\tilde{\alpha}_{\mathrm{Band}}$ & 68\% interval \\
   
    &&&&\\
    
         LGRB w/ break & 12& $-0.75$ & $-0.76$ & $[-0.90,-0.57]$ \\
         LGRB w/o break & 15 & $-0.94$ & $-0.95$& $[-1.08,-0.80]$ \\
         SGRB & 9&$-0.38$ & $-0.47$ & $[-1.03,0.3]$ \\
    \hline
         
    \end{tabular}
    \caption{Characteristic values (mean, median and 68\% interval) of the distributions of $\ab$ shown in the top panel of Fig.~\ref{fig:a1_a2_ab_dist}. The  number of GRBs in each sample is reported in column 2.}
    \label{tab:hist_top}
\end{table}

\begin{table}[htbp]
\centering
    \begin{tabular}{c|c|c|c}
    \hline
    \hline
          Index &  $\langle \alpha_{i} \rangle$ &$\tilde{\alpha}_{i}$ & 68\% interval \\
&&&\\

         $\au$ & $-0.71$ &$-0.70$&$[-0.86,-0.60]$\\
         $\ad$ & $-1.71$ & $-1.69$& $[-1.82,-1.62]$ \\
    \hline
    \end{tabular}
    \caption{Characteristic values (mean, median and 68\% interval) of the distributions of  $\au$ and $\ad$ shown in the bottom panel of Fig.~\ref{fig:a1_a2_ab_dist} for the 12 LGRBs best fitted with the 2SBPL.}
    \label{tab:hist_bottom}
\end{table}

In the bottom panel of Fig.~\ref{fig:a1_a2_ab_dist} we show the distributions of 
the indices $\au$ (red) and $\ad$ (violet) for the 12 LGRBs best fitted by the 2SBPL (i.e. with identified low-energy spectral break). 
The characteristic values (mean, median and 1$\sigma$ dispersion) of the distributions in Fig.~\ref{fig:a1_a2_ab_dist} are reported in Table~\ref{tab:hist_top} and \ref{tab:hist_bottom}.

From the comparison of the distributions shown in Fig.~\ref{fig:a1_a2_ab_dist} we find that: 
\begin{enumerate}

    \item SGRBs (green dashed histogram) have a harder spectral slope $\ab$ than LGRBs without a break (blue histogram).  A Kolmogorov-Smirnov (KS) test\footnote{For all the  statistical
    tests we have set the significance level at 0.05, i.e. we accept the null 
    hypothesis if $p>0.05$.} among the two distributions returns a $p$-value of 0.004, rejecting the null hypothesis of being drawn from the same underlying distribution. This is consistent with previous studies \citep{ghirlanda04,ghirlanda09};

    \item the value of $\ab$ for LGRBs with a 
    break (orange histogram in Fig.~\ref{fig:a1_a2_ab_dist}, top panel) is on average harder (see Table~\ref{tab:hist_top}) than the value for LGRBs with no break (blue histogram). However, the two distributions are not distinguishable (a KS test between the orange and blue distributions has a 
    chance probability $p=0.08$);  

    \item the distributions of $\au$ and $\ad$ (red and violet histograms in  Fig.~\ref{fig:a1_a2_ab_dist}, bottom panel) are peaked at --0.71 and --1.71, not far from the  typical  values --2/3 and --3/2 expected for synchrotron spectrum from marginally fast cooling 
    electrons;

    \item LGRBs without a break have an $\ab$ distribution slightly softer than $\au$ but harder than $\ad$
    (cfr. the blue histogram in the top panel with the red and violet histogram in the bottom panel of Fig.~\ref{fig:a1_a2_ab_dist} respectively). This might suggest that when the spectral data are not sufficient to constrain and identify a spectral break, the Band function returns a value of the low-energy index that is an average between the index $\au$ and $\ad$. This possibility will be investigated with simulations in the next section;

    \item the distribution of $\ab$ of SGRBs (green) is similar to the $\au$ 
    distribution of LGRBs with break: a KS test between the two returns $p=0.16$. This suggests that the power-law segment $\ad$ separating $\eb$ from $\ep$ is not present in SGRBs, i.e. $\eb\sim\ep$.
\end{enumerate}

\section{Origin of the value $\alpha_{\mathrm{Band}}\sim -1$}

The spectral analysis performed in this paper confirmed the presence of two classes of LGRBs: those requiring two power-law segments ($\au$ and $\ad$) to describe the spectrum at energies $E < E_{\rm peak}$, and those for which this part of the spectrum is well described by a single power-law ($\ab$). Since the values of $\ab$ are typically softer than $\au$ but harder than $\ad$, we investigate the possibility that spectra best fitted by Band are hiding a spectral break that is difficult to identify with a certain statistical significance due to the lack of enough signal at low energies, and/or to the proximity of $\eb$ to $\ep$, and/or to the proximity of $\eb$ to the low energy edge of the GBM sensitivity. If this is correct, we would expect to see a dependence of $\ab$ on the values of $\eb$ and $\ep$ and on their separation.
Specifically, we expect that when the underlying spectrum has a break, the fit with the Band function would return a hard $\ab\sim\au$ when $\eb\sim\ep$, and, conversely, a soft $\ab\sim\ad$ when $\eb \ll \ep$.

A strong correlation is not expected, as the value of $\ab$ should depend not only on the ratio $\er = \eb/\ep$, but also on the absolute value of $\ep$ (or, equivalently, $\eb$), and also on the specific values of $\au$ and $\ad$.
To better investigate this effect and its presence in the spectra, we performed a set of simulations that are described in the following sections. 

\subsection{Band function response to a spectral break}\label{sec:band_trend}

In this first subsection we investigate how the presence of a spectral break generally affects the results of a fit performed using the Band function.
We simulate GRB prompt spectra with input model 2SBPL, keeping fixed all the parameters  and varying solely $\eb$. 
The adopted input parameters are $\au=-0.65$, $\ad=-1.67$, $\ep=1000$ keV, $\bs= -2.5$.
These input values have been chosen in order to reproduce a typical LGRB of our sample (see the fit results in $\S3$).
 For these simulations, we use the GBM background and response matrix files from one of the GRBs  in our sample. We verified that choosing different background and response matrix files belonging to any other GRB in our sample does not affect the simulation results.

\begin{figure*}
    \centering
    \includegraphics[scale=0.6]{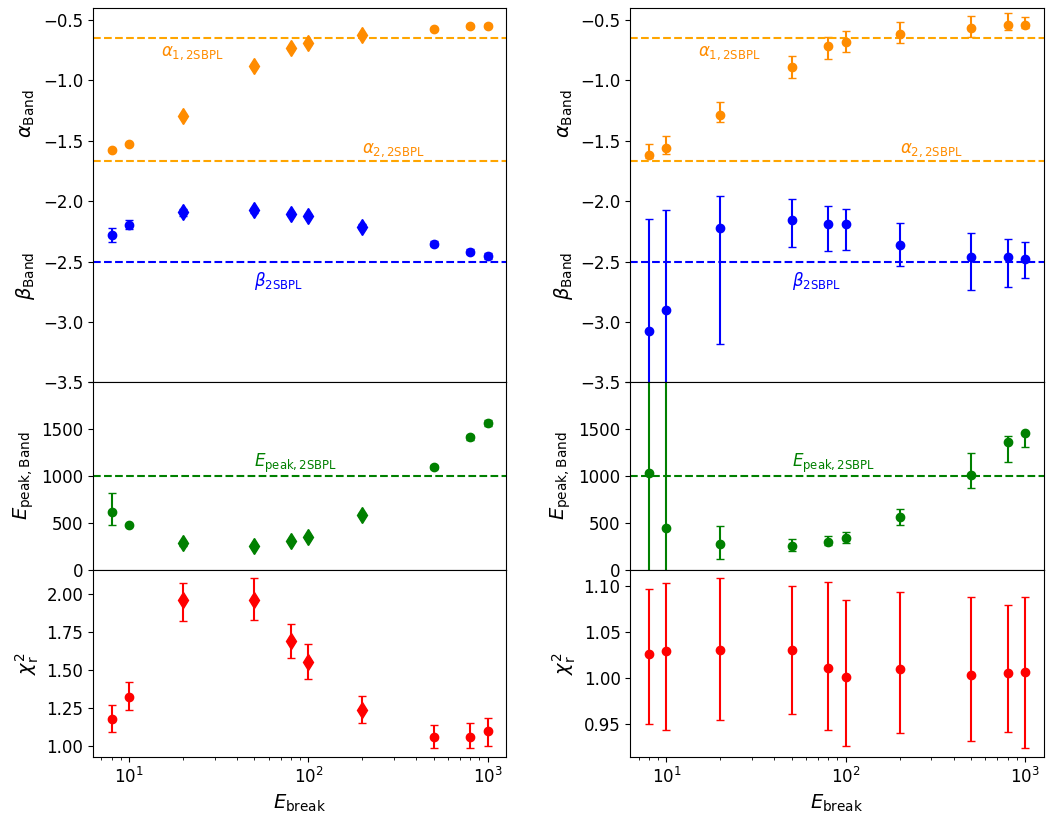}
    \caption{Band function parameters  as a function of the position of the energy break $\eb$ of the 2SBPL function. Each plot shows the parameters of the Band function fitted to a series of spectra simulated assuming the 2SBPL function whose parameter values are marked by the horizontal dashed lines. 
    Top: low-energy slope $\ab$ (orange symbols) and high-energy slope $\bb$ (blue symbols); middle: energy peak $\epb$ (green symbols); bottom: fit $\chi_\mathrm{r}^2$ (red symbols).
    Left: spectrum characterized by a S/N $\sim 21$ ( fluence $\sim 3.5\cdot 10^{-4}$ erg cm$^{-2}$); right: spectrum characterized by S/N $\sim 2.7$ ( fluence $\sim3.5\cdot 10^{-5}$ erg cm$^{-2}$).
    Data is represented as circles when the best fitting model is Band; as diamonds when the best fitting model is 2SBPL.}
    \label{fig:band_trend}
\end{figure*}

Each simulated spectrum is then fitted with the input model (a 2SBPL with parameters free to vary) and also with a Band function. 
For each value of $\eb$, we repeated the simulation 200 times, obtaining (for each parameter and for the reduced chi-square) a distribution of values. From these distributions we extracted the mean value and its 68\% confidence interval.
Figure \ref{fig:band_trend} shows the parameters returned by the Band fits as a function of the position of the energy break.
This exercise is repeated for two different cases, with a rather high average S/N ratio\footnote{  calculated as $(s - b)/\sqrt{b}$, where $s$ and $b$ are the source and background estimated counts, respectively (see e.g. \citealt{dereli20}). ($\sim 21$, left-hand panel) and a S/N ratio that is approximately a factor 10 lower ($\sim 2.7$, right-hand panel).
They represent  simulated spectra of a GRB with a fluence of $\sim 3.5\cdot 10^{-4}$ erg cm$^{-2}$ and $\sim 3.5\cdot 10^{-5}$ erg cm$^{-2}$, respectively.}
The input parameters used for the  2SBPL function used for the simulations 
($\au$,$\ad$,$\ep$, $\bs$) are marked by dashed horizontal lines. 
We distinguish the best fitting model according to our criterion based on the $\mathrm{AIC}$ (in Fig.~\ref{fig:band_trend}, diamonds: 2SBPL, circles: Band).

The values of $\ab$ obtained fitting
the simulated spectra with the Band function (orange symbols in the top panel)  
correlate with $\eb$: a low value of $\eb$ makes $\ab\approx\ad$. 
On the other hand, as $\eb$ increases (and approaches $\ep$ which in this 
example is 1 MeV) 
$\ab\approx\au$. In between, the value 
of $\ab$ is an average of $\au$ and $\ad$, depending on the position of $\eb$.
Given the presence of only a single break in the Band function (i.e. $\epb$) also the other parameters ($\bb$ and $\epb$) depend on the position of the break: $\bb$ (blue symbols) always assumes softer values compared to the input one, unless  $\eb \sim \ep$. $\epb$ (green symbols) is an average of $\eb$ and $\ep$ of the 2SBPL function and approaches the input value when $\eb$ is very low or when $\eb\sim\ep$. 

These results hold for both cases of the S/N.
The main difference is in the uncertainties on the best fit parameters (larger for the case with lower  S/N) and, most notably, on the behavior of the $\chi_\mathrm{r}^2$.
In the case with lower  S/N, the $\chi_\mathrm{r}^2$ of the Band fit is always acceptable ($\sim 1$), no matter the value of $\eb$. 
This shows that, even though the input spectrum has a spectral break and this break falls within the GBM energy range, in a spectrum with a relatively low  S/N
the identification of the break is not possible, and the best fit model is a Band function. 
Note that a fluence of $3.5\cdot 10^{-5}$ erg cm$^{-2}$ or less is representative of the majority of LGRBs detected by {\it Fermi}/GBM.
If the  S/N is increased by a factor of 10 (left-hand panel) the $\chi_\mathrm{r}^2$ of the fit with the Band function depends on $\eb$: only when the break is at the very low energy end of the GBM spectral range ($\eb\lesssim10$ keV)  or close to $\ep$  ($\eb\gtrsim 500$ keV) the fit with the Band function returns an acceptable $\chi_\mathrm{r}^2$. Despite the high  S/N, in such cases the break is hardly identifiable ($\Delta \mathrm{AIC}<6)$.

Finally, we notice that even when the Band function returns an adequate fit, (i.e. when $\eb\lesssim$ 10\,keV or $\eb\sim\ep$) the resulting values of $\ab$, 
$\bb$ and $\epb$ might largely deviate from the values of the input spectrum.

\subsection{Spectral simulations: $\er - \ab$ trend} \label{sec:simulations}

\begin{figure*}[htbp]
    \centering
    \hspace*{-0.6cm}
    \includegraphics[scale=0.62]{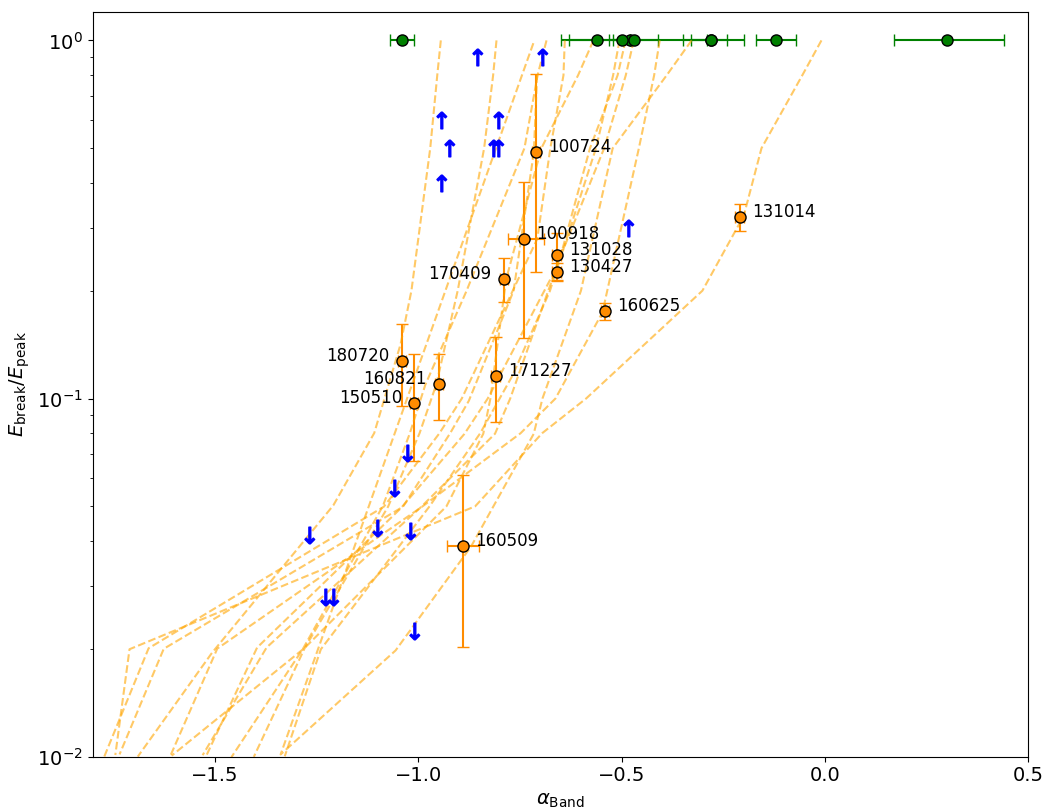}
   
    \caption{The twelve LGRBs best fitted by a 2SBPL (Table~\ref{tab:fit_params_w/_break}) are shown with 
     orange symbols. Their value of $\er=\eb/\ep$ is shown vs. the value of the index $\ab$ that is obtained by fitting their spectrum with a Band function.
    SGRBs (whose spectrum is always best fitted by the Band function) are here represented
    assuming that, if the underlying spectrum were 2SBPL, it should have $\eb\sim\ep$ (green symbols).
    The orange dashed lines show the results of simulations (see \S 4.2);
       blue arrows represent the upper and lower limits on $\er$ for the 15 LGRBs whose spectra 
    are best fitted by Band (see \S 4.3). }

    \label{fig:alpha_plot}
\end{figure*}

In order to further investigate the $\er -\ab$ trend, we focus first on the 12 LGRBs analyzed in this work that have a spectral break $\eb$.
In Fig.~\ref{fig:alpha_plot} we show (orange symbols) their ratio $\er=\eb/\ep$ (from the fits 
of the 2SBPL) versus $\ab$ (from the fit of the same spectrum with the Band function).
LGRBs with a break are located in the range $\er\in[0.04,0.5]$ and, if 
their spectra are fitted with the Band function, the resulting $\ab$ is in the range $\ab\in[-1.1, -0.2]$. 
A broad trend in the $\er-\ab$ plane appears among the points. 
The Pearson correlation coefficient is 0.56 and the associated chance probability value $p=0.05$.

For each of these GRBs, we simulate\footnote{Spectral simulation 
performed within \texttt{XSPEC} with the \texttt{fakeit} tool.} 
spectra with the 2SBPL 
function with parameter values fixed to the best fit values (reported in Table~\ref{tab:fit_params_w/_break})
except for $\eb$, that we vary between 0.01$\ep$ and $\ep$. 
 For each GRB, we used the corresponding GBM background and response matrix files for the corresponding simulations.
Simulated spectra were renormalized in order to maintain the energy--integrated flux 
of the real spectrum constant while moving $\eb$.  
Low values of $\er$ places the break below the GBM low energy 
threshold, i.e. 8 keV, in those GRBs with $\eps<800$\,keV. 
We then refit the simulated spectra with the Band function and derive $\ab$. The simulation of each spectrum is repeated 200 times to build the distribution of $\ab$ and estimate its mean value and 68\% confidence interval.

In Fig.~\ref{fig:alpha_plot} we show, for each one of the 12 LGRBs, the corresponding $\ab$ returned by the fit with the Band function for each input value of $\er$ (orange dashed line). 
These curves show that $\ab$ depends on the relative position between break and peak energy, with small ratios resulting in soft spectra and large ratios resulting in harder spectra, as expected.
These simulations show that the value of $\ab$ is a ``weighted'' mean of the $\au$ 
and $\ad$ slopes.
Moreover, different curves show similar trend, showing that the different values of $\ep$ and input $\au$ and $\ad$ are responsible for the dispersion in the plane. The dispersion of the curves is similar to the dispersion in the real data.

From the tracks of the orange dashed lines shown in Fig.~\ref{fig:alpha_plot} we can speculate that, when the best fit model is a Band function, values of $\ab$ harder than $\sim -1.0$ could be consistent with the presence of an $\eb$  in the proximity 
(i.e. till one order of magnitude lower) of $\ep$, while values softer 
than $\sim -1.0$, could indicate the presence of $\eb$ far from (i.e. more than one order 
of magnitude lower than) $\ep$.
This possibility is investigated in the next section, through spectral simulations.

\subsection{Spectral simulations: origin of the spectra without a low-energy break}

Fifteen LGRBs in our sample do not show the presence of a low-energy spectral break. 
Through spectral simulations, we now propose to investigate if it is still possible for these GRBs to have a low energy break, even though the best fit model is a simple Band function. 
The simulation now assumes that also in these 15 GRBs the spectrum has 
the shape of the 2SBPL function and infers constraints on its parameters by requiring that the fit with the Band function is not only acceptable but also preferred over a 2SBPL, and returns as best fit values the same values as the real spectrum.

Based on the trend found between $\er$ and $\ab$,  we would expect that 
if the spectrum is intrinsically a 2SBPL and $\eb$ lies at low energies, 
the Band function could adequately fit the simulated spectrum and result in $\ab\sim\ad$.
Similarly, if the spectrum is intrinsically a 2SBPL and $\eb$ lies close to $\ep$, 
the Band function could return  $\ab\sim\au$. 
For each burst the  spectra simulated with the 2SBPL function maintain the same fluence of the real GRB. 

We repeat the simulations for different values of the 2SBPL parameters. In particular: 

\begin{itemize}

    \item $\au$ is sampled uniformly within the range $\au \in [-0.6,-0.87]$\footnote{This interval corresponds to the 68\% interval obtained from the analysis of 
    the 12 LGRBs with identified break (Table~\ref{tab:hist_bottom})} with steps of 0.03; 
    
    \item $\ad$ is sampled uniformly in the interval $\in[-1.1,-1.9]$ with steps of 0.03;
    
    \item $\eb$ is sampled between 2 and 50 keV with steps of 2 keV;
    
    \item $\bs$ is fixed to the value obtained from the fit with the Band function.
\end{itemize}

For each combination of parameters, we simulate 10 spectra.  We assume the background and response matrix files of each GRB for these simulations. 
They are then refitted with both the 2SBPL and Band functions. 
From the built parameter distributions we derive the mean values and 68\% confidence interval.
Once we refit the spectrum with a Band function we accept the simulation if the Band fit  satisfies the following conditions: 
\begin{itemize}

    \item it is statistically equivalent to the fit with the 2SBPL, i.e. 
    $\Delta \mathrm{AIC} <6$;

    \item its $\ab$ and $\bb$ are consistent, within 1$\sigma$, with the values inferred from the real spectrum;

    \item its $\ep$ is consistent,  within 3$\sigma$, with the value inferred from the real spectrum.
\end{itemize}

For each of the 15 LGRBs which do not explicitly show a break, we find a significant number 
of parameters' combinations for which the 2SBPL functions were able to satisfactorily reproduce the real spectrum.
 
In particular, for all these LGRBs we are able to  set  either a plausible maximum or minimum  value for $\eb$ and constrain either $\ad$ or $\au$ in an interval.
These are represented with the blue arrows in Fig.~\ref{fig:alpha_plot}. 
The limits for $\eb$ and the low-energy slopes intervals are listed in Table~\ref{tab:constraints_down} and Table~\ref{tab:constraints_up}. 

\begin{table}[htbp]
    \centering
    \begin{tabular}{c|c|c}
    \hline
    \hline
   Name      &     $\ad$ Range    &       $E_{\mathrm{break,Max}}$   \\ 
   &&\\
090323(002)    &      $[-1.48, -1.18]  $   &          30            \\ 
130504(978)  &      $[-1.42, -1.28]$    &          12\\
130606(497)    &   $[-1.24, -1.12] $       &       12\\
140206(275)     &     $[-1.60, -1.36]$     &          18\\
170210(116)     &     $[-1.37, -1.14]$     &         24\\
170527(480)      &    $[-1.34, -1.15]$     &         24\\
\hline

    \end{tabular}
    \caption{Constraints on $\ad$ and maximum  $\eb$ (in keV) for  GRBs with soft $\ab$ which did not show an energy break in their time-integrated spectrum. }
    \label{tab:constraints_down}
\end{table}

\begin{table}[htbp]
    \centering
    \begin{tabular}{c|c|c}
    \hline
    \hline

 Name      &     $\au$ Range    &     $E_{\mathrm{break,Min}}$ \\ 
   &&\\
090926(181)    &      $[-0.83, -0.80]  $   &    220   \\    
100414(097)  & $[-0.68, -0.54]$       &         126     \\ 
101123(952)    &  $[-1.02, -0.94]$       &       182         \\ 
120526(303)   &   $[-0.90, -0.83]$         &        372      \\ 
120624(933)    & $[-1.05, -0.93]$     &            320     \\  
120711(115)  &       $-0.95$                 &      440      \\
130306(991)  &         $[-0.98,-0.72]$  &           86      \\ 
160905(471)  &       $[-0.90,-0.84]$         &    572        \\ 
170214(649)  &        $[-0.92,-0.86]$        &   157\\        
\hline

    \end{tabular}
    \caption{Constraints on $\au$ and minimum  $\eb$ (in keV) for  GRBs with hard $\ab$ 
    which did not show an energy break in their time-integrated spectrum. }
    \label{tab:constraints_up}
\end{table}

\section{Discussion and Conclusions}

The prompt emission spectra of long GRBs are often fitted with the Band function, two power laws smoothly joined at the $\nu F_{\nu}$ peak. The low energy index (below the peak energy $\ep$), $\ab\sim -1$ has been used as an argument against the interpretation of the prompt emission as synchrotron (see e.g. \citealt{preece98,frontera00,ghirlanda02}).  Recently different groups identified a break, $\eb$, at low energies below $\ep$ \citep{oganesyan17,oganesyan18,oganesyan19,ravasio18,ravasio19} paving the route towards the solution of the long-standing issue on the nature of the prompt emission process (see e.g.  \citealt{daigne11,uhm14,bosnjak09,ghisellini99,rees94,sari96,sari98}).  

According to these works, the prompt emission spectra of the brightest GRBs can be described with three power laws (with indexes $\au$ below $\eb$, $\ad$ between $\eb$ and $\ep$ and $\beta$ above it) smoothly joined at the two breaks, namely  $\eb$ and $\ep$.

If the spectrum is a 2SBPL, our simulations described in \S4.1 show that when $\eb$ is close to $\ep$ or below the low energy threshold ($E_{\mathrm{min}}$) of the instrument, the Band function gives $\ab\sim\ad$ and $\ab \sim \au$, respectively. Values of $\ab\sim-1$ correspond to $\eb$ between $E_{\mathrm{min}}$ and $\ep$.  Through the spectral analysis of sample of GRBs, selected with different criteria, \cite{burgess20} find that, when $\eb\lesssim\ep$, the values of $\ab$ are distributed approximately $\in[-1.7,-0.5]$.
 We argue that, if the break is a common feature of GRB spectra, the value of $\ab$ is a proxy of its position with respect to $\ep$. 
 
\begin{figure}[htbp]
    \centering
    \includegraphics[scale=0.5]{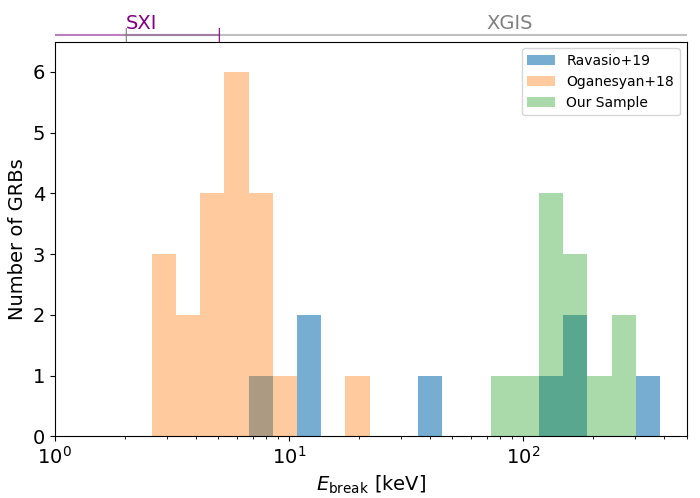}
    \caption{Energy distribution of spectral breaks identified in prompt spectra of GRBs, from the results of \cite{oganesyan18}, \cite{ravasio19} and this work. 
    The horizontal lines indicate the observational energy range of the instrumentation aboard THESEUS.}
    \label{fig:known_break_dist}
\end{figure}

This hypothesis is verified through the spectral analysis of a sample of 27 long and 9 short GRBs selected, within the {\it Fermi} sample, with large fluence and large $\ep$ (\S2) in order to ease the search of $\eb$, if present. In 12/27 long GRBs we find $\eb$ (i.e., the 2SBPL better fits the data with respect to Band). Using these events as templates, through spectral simulations, we find that if the break is moved within the range delimited by $\ep$ and $E_{\mathrm{min}}$,  the fit with Band results in a softer (if $\eb$ departs from $\ep$) or harder (if $\eb$ approaches $\ep$) low energy index $\ab$ (dashed orange lines in Fig. \ref{fig:alpha_plot}). At the extreme the values of $\au$ and $\ad$ are found. Indeed, none of the short GRBs analyzed has a break but  a relatively hard $\ab$ which we suggest corresponds to $\eb$ lying close to $\ep$. Through dedicated spectral simulations (\S4.3) we show that the 15 long GRBs best fitted by the Band function only (i.e. apparently without a break) could instead have a break close to $\ep$, corresponding to $\ab>-1$ (upward arrows in Fig. \ref{fig:alpha_plot}), or close to $E_{\mathrm{min}}$  if $\ab<-1$ (downward arrows in Fig. \ref{fig:alpha_plot}).

Our analysis suggests that the low energy break could be a feature more common than what can be inferred by direct spectral analysis. Indeed, the identification of the break in the spectra of GRBs detected by {\it Fermi} or {\it Swift}, now possible only for a limited number of events (shown in Fig. \ref{fig:known_break_dist}), is hampered by 1) the separation of $\eb$ from $\ep$ and 2) the spectral signal-to-noise ratio. 
We have shown (right panel of Fig. \ref{fig:band_trend}) that a burst with a typical fluence (e.g. $5\cdot 10^{-6}\, \mathrm{erg \, cm^{-2}}$  ) detected by {\it Fermi}/GBM can be fitted by Band even if it has an additional break.  These effects concur to explain why in approximately half of the selected GRBs we could not find $\eb$ but, through simulations, we could set an upper/lower limit to its possible value. 

With the currently available instruments, {\it Swift} and {\it Fermi}, it was possible to find $\eb$ in a limited number of GRBs and with $\eb$ at X--ray  ($\sim$ few keV) and $\gamma$--ray ($\sim$ tens - hundreds keV) energies (Fig. \ref{fig:known_break_dist}). A few values of $\eb$ between 10 and 100 keV are found. 

 Our results (Fig. \ref{fig:a1_a2_ab_dist} - bottom panel) show that the distributions of $\au$ and $\ad$ are close to but slightly softer than the values predicted by synchrotron emission in moderate fast cooling regime \citep{daigne11}, i.e. -3/2 and -2/3, respectively. This is partly due to our fits with the 2SBPL function rather than with the synchrotron model (see e.g. \citealt{burgess15,burgess20}) and to the fact that we analyze time integrated spectra to exploit the largest S/N ratio in search of $\eb$. Time resolved spectral analysis, indeed, often finds harder spectral slopes \citep{nava11a, acuner18} and, as shown by \cite{ravasio19}, the distributions of $\au$ and $\ad$ are closer to the typical synchrotron values.

With the Transient High-Energy Sky and Early Universe Surveyor (THESEUS) mission \citep{amati18,amati2021} proposed to ESA within the M5-class 
selection call, we expect that the spectral break will be detected in a 
larger fraction of events \citep{ghirlanda2021}. 
The large effective area and the wide energy range covered by the two 
instruments on board THESEUS, Soft X-ray Imager (SXI, 0.3-5 keV) and X-Gamma rays Imaging Spectrometer (XGIS, 2 keV--few MeV), 
will provide us prompt emission spectra with high statistics from which 
$\eb$ will be measured over a wider fluence range than what is now possible.

\begin{acknowledgements}
GO acknowledges financial contribution from the agreement ASI-INAF n.2017-14-H.0. G. Ghirlanda acknowledges the Premiale project FIGARO 1.05.06.13 and INAF-PRIN 1.05.01.88.06.

\end{acknowledgements}

\bibliographystyle{aa}
\bibliography{references}
\end{document}